\documentclass[letterpaper]{ieeeconf}  
\usepackage{authblk}
\usepackage{cite} 
\usepackage{booktabs,tabularx}

\usepackage{graphicx} 
\usepackage{amsmath}
\usepackage{array}    
\usepackage{caption}
\usepackage{hyperref}

\title{
Emulating Clinical Quality Muscle B-mode Ultrasound Images from Plane Wave Images Using a Two-Stage Machine Learning Model
}

\author[1]{Reed Chen}
\author[1]{Courtney Trutna Paley}
\author[1]{Wren Wightman}
\author[2]{Lisa Hobson-Webb}
\author[2]{Yohei Harada}
\author[2]{Felix Jin}
\author[1]{Ouwen Huang}
\author[1]{Mark Palmeri}
\author[1]{Kathryn Nightingale}

\affil[1]{Duke University}
\affil[2]{Duke University Medical Center}

\begin{document}

\maketitle

\begin{abstract}
Research ultrasound scanners such as the Verasonics Vantage™ often lack the advanced image processing algorithms used by clinical systems. Image quality is even lower in plane wave imaging -- often used for shear wave elasticity imaging (SWEI) -- which sacrifices spatial resolution for temporal resolution. As a result, delay-and-summed images acquired from SWEI have limited interpretability. In this project, a two-stage machine learning model was trained to enhance single plane wave images of muscle acquired with a Verasonics Vantage™ system. The first stage of the model consists of a U-Net trained to emulate plane wave compounding, histogram matching, and unsharp masking using paired images. The second stage consists of a CycleGAN trained to emulate clinical muscle B-modes using unpaired images. This two-stage model was implemented on the Verasonics Vantage™ research ultrasound scanner, and its ability to provide high-speed image formation at a frame rate of 28.5 ± 0.6 FPS from a single plane wave transmit was demonstrated. A reader study with two physicians demonstrated that these processed images had significantly greater structural fidelity and less speckle than the original plane wave images. 

\end{abstract}

\section{Introduction}

Ultrasound is a widely used medical imaging modality due to it being low-cost, portable, and non-ionizing. However, ultrasound images are degraded by speckle, electronic noise, differences in tissue sound speed, and other sources of error \cite{RN1, RN2, RN3}. Because of this, commercial clinical ultrasound scanners use proprietary post-processing algorithms to enhance images \cite{RN4}. Unfortunately, research scanners, such as the Verasonics Vantage™, lack these post-processing algorithms, often resulting in images that appear to be of lower quality. Because reader studies have demonstrated that clinicians prefer the post-processed images that clinical systems offer \cite{RN5}, it is difficult to accurately compare images from research scanners to images from clinical ultrasound systems.

Previous studies have attempted to address this disparity between images from research scanners and clinical scanners using machine-learning approaches. Huang et. al. introduced MimickNet, a CycleGAN that converts conventional delay-and-summed images into approximate Siemens Dynamic Tissue Contrast Enhanced (DTCE™) post-processed images \cite{RN4}. The utility of MimickNet was successfully demonstrated on fetal, phantom, and liver targets. Additionally, histogram matching has also been proposed as a normalization method to fairly compare images across ultrasound image formation methods \cite{RN6}. In other studies \cite{RN7, RN8, RN9, RN10}, researchers used CycleGANs to approximate images acquired with cart-based ultrasound systems from images acquired with point-of-care ultrasound devices for diverse tissues including cardiac, thyroid, carotid, breast, and forearm and calf muscle. In \cite{RN11}, authors applied a GAN to translate ultrasound images acquired with diverse imaging systems and parameters into images that were seemingly taken from the same system. 

Another disadvantage of focused delay-and-sum ultrasound (fDAS-US) is the time required to acquire each frame. For a B-mode image to be produced, multiple focused transmit beams are used, limiting the frame rate. This makes fDAS-US less unsuitable for applications such as shear wave imaging or elastography where high frame rates are required \cite{RN12}. For these applications, plane wave imaging and plane wave compounding are often used instead. Plane wave imaging uses a single unfocused plane wave transmit to insonify the tissue in the field of view. Because the frame rate is only limited by the duration of a single round-trip pulse-echo, kilohertz frame rates can be achieved using plane wave imaging. However, because the transmit beam is unfocused in plane wave imaging, the main lobe of the ultrasound system’s point spread function is larger, reducing image resolution. Additionally, the total acoustic output is lower with fewer transmits, resulting in decreased SNR \cite{RN13}.

To compensate for the limitations of plane wave imaging, plane wave compounding is often utilized. In plane wave compounding, multiple plane waves transmitted at different angles are coherently compounded to acquire improved images \cite{RN12}. Although spatial resolution and contrast improve as the number of plane waves transmitted increases, frame rate decreases. Because of this decrease in frame rate, past work has demonstrated that it is possible to emulate plane wave compounded (PWC) images from a single plane wave using machine learning approaches \cite{RN14, RN15, RN16}. There has also been research in applying CNNs, GANs, and CycleGANs for the enhancement and super-resolution of single plane wave images of phantoms, carotid, and thyroid \cite{RN17, RN18, RN19}.

\begin{table*}[t]
    \caption{Clinical muscle ultrasound repositories compiled to train the CycleGAN}
    \label{tab:1}
    \begin{tabularx}{\textwidth}{c*{2}{>{\raggedright\arraybackslash}X}cc}
        \toprule
        \textbf{Dataset} & \textbf{Muscle Description} & \textbf{Scanner System} & \textbf{Images} & \textbf{Images after Cleaning} \\
        \midrule
        1 \cite{RN26} & Longitudinal images of muscle-tendon junctions on the lateral and medial gastrocnemius. & Aixplorer V6 (SL10-2, 9 MHz), Esaote MyLab60 (LA923, 7 MHz), and Telemed ArtUs (LV8-5N60-A2, 8 MHz). & 1344 & 505 \\
        \midrule
        2 \cite{RN27, RN28} & Longitudinal images of the lateral and medial gastrocnemius, vastus lateralis, tibialis anterior, and soleus. & AlokaSSD-5000 PHD (7.5 MHz) and other ultrasound devices. & 883 & 504 \\
        \midrule
        3 \cite{RN29} & Majority transverse images of the biceps brachii, tibialis anterior, and medial gastrocnemius. & Esaote MyLab Twice (3–13 MHz). & 3917 & 2354 \\
        \midrule
        4 \cite{RN30} & Longitudinal images of the vastus lateralis and lateral and medial gastrocnemius. & GE Logiq E9 (11 MHz), Telemed LogicScan 128 EXT-1Z (LV7.5/60/96Z), Telemed SmartUs (LV7.5/60/128Z-2), and Philips HD11XE (5-12 MHz). & 339 & 137 \\
        \bottomrule
    \end{tabularx}
\end{table*}

One important application for ultrasound is muscle imaging. Muscle ultrasound is often used for diagnosing myopathies and muscle injuries \cite{RN20}. There has also been a large interest in musculoskeletal SWEI as studies have indicated a link between shear wave speeds and myopathies, dystrophies, and muscle spasticities \cite{RN20, RN21, RN22}. However, muscle is a complex tissue due to the presence of transversely isotropic muscle fibers and fascicles. In transverse images where the transducer is perpendicular to the direction of the muscle fibers, muscle has an appearance dubbed the “starry night appearance” \cite{RN23}. In longitudinal images where the transducer is parallel to the direction of the muscle fibers, the muscle fascicles appear as relatively uniform parallel lines. This dependence on transducer orientation poses a challenge for post-processing algorithms designed for muscle. 

\begin{figure}[ht]
    \centering
    \includegraphics[width=0.5\textwidth]{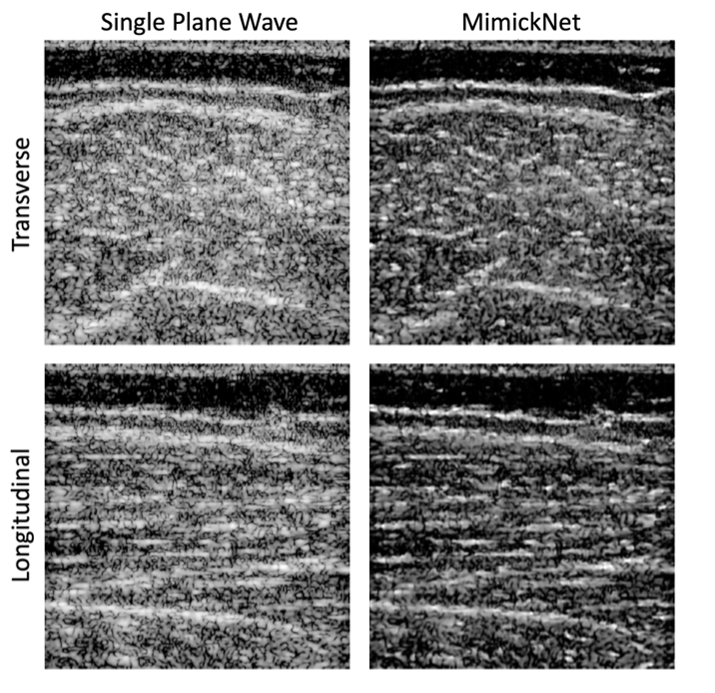}
    \caption{Left column: Single plane wave images taken along the transverse and longitudinal views of the vastus lateralis. Right column: Single plane wave images after processing by MimickNet.}
    \label{fig:1}
\end{figure}

Our group has observed that SWEI DAS images, due to their low resolution, have limited utility for identifying tissue structures or further analyses such as tissue segmentation. Currently, there is a lack of post-processing algorithms that can enhance plane wave images of muscle which are often acquired for muscle SWEI in real-time. As seen in Figure \ref{fig:1}, applying MimickNet to muscle images yielded poor-quality images as the model was not trained on muscle data nor plane wave data, rather B-mode images of fetal, phantom, and liver targets \cite{RN4}. To address this gap, we present a two-stage machine-learning approach to approximate clinical quality images of muscle from single plane waves. These machine learning models are implemented on the Verasonics Vantage™ system to provide real-time enhancement of single plane wave muscle images.

\section{Methods}
To train each stage of the two-stage machine learning model independently, two datasets were used: a paired dataset for the first stage, and an unpaired dataset for the second stage.

\subsection{Paired Dataset}
The paired dataset consists of 15,948 images of the vastus lateralis muscle (4 cm lateral by 4 cm axial) acquired from 10 healthy volunteers under a Duke IRB-approved protocol. These images were collected on the Verasonics Vantage™ using an L7-4 probe during SWEI acquisitions. The setup for collecting these images is described by Paley et. al. \cite{RN24}, and consists of placing the transducer 2/3 the distance from the lateral condyle to the greater trochanter of the femur of each volunteer’s self-described dominant leg.  Because muscle is transversely isotropic due to the presence of muscle fibers, it was necessary to train the machine learning model on both transverse (across the fiber) and longitudinal (along the fiber) images of muscle. The transducer was rotated by a Newport rotation stage across 36 different angles spanning 180 degrees, with the transducer repositioned as needed to maintain contact during the rotation. Thus, the dataset includes both transverse and isotropic views of the vastus lateralis. In this paper, we consider any images taken at the 18 angles most parallel to the muscle fibers to be longitudinal images, and images taken at the remaining 18 angles are considered transverse. 

The beamformed data from a single non-steered (0°) plane wave was envelope-detected and log-compressed to generate the input training images. Paired ground truth images were generated using 12 plane waves acquired at 3 different angles: -3°, 0°, and 3°. These 12-steered plane wave images were coherently compounded \cite{RN12}, envelope-detected, and log-compressed. These images were then filtered with a traditional image processing pipeline consisting of histogram matching and unsharp masking \cite{RN6,RN25} to produce the final ground truth images. The SWEI B-modes had an image size of 97x191 pixels before being upsampled to 512x512 pixels using bicubic interpolation.

\subsection{Unpaired Dataset}
The second stage of the model is a CycleGAN trained to perform style transfer between muscle images acquired with the Verasonics Vantage™ research scanner and clinical ultrasound scanners. The input image domain consists of the compounded and filtered images as described in the Paired Dataset section. The dataset used for the clinical image domain consists of images sourced from multiple online repositories listed in Table \ref{tab:1}.

Images containing artifacts from dead elements, poor contact, and motion were excluded. Images that had very low resolution, were blurry, or contained substantial annotations were also removed. Images that contained annotations located in the periphery of the image or were screenshots were cropped to isolate the ultrasound image. The total number of images in the clinical domain dataset decreased from 6,483 to 3,500 after removing these poor-quality images.

\begin{figure}[ht]
    \centering
    \includegraphics[width=0.5\textwidth]{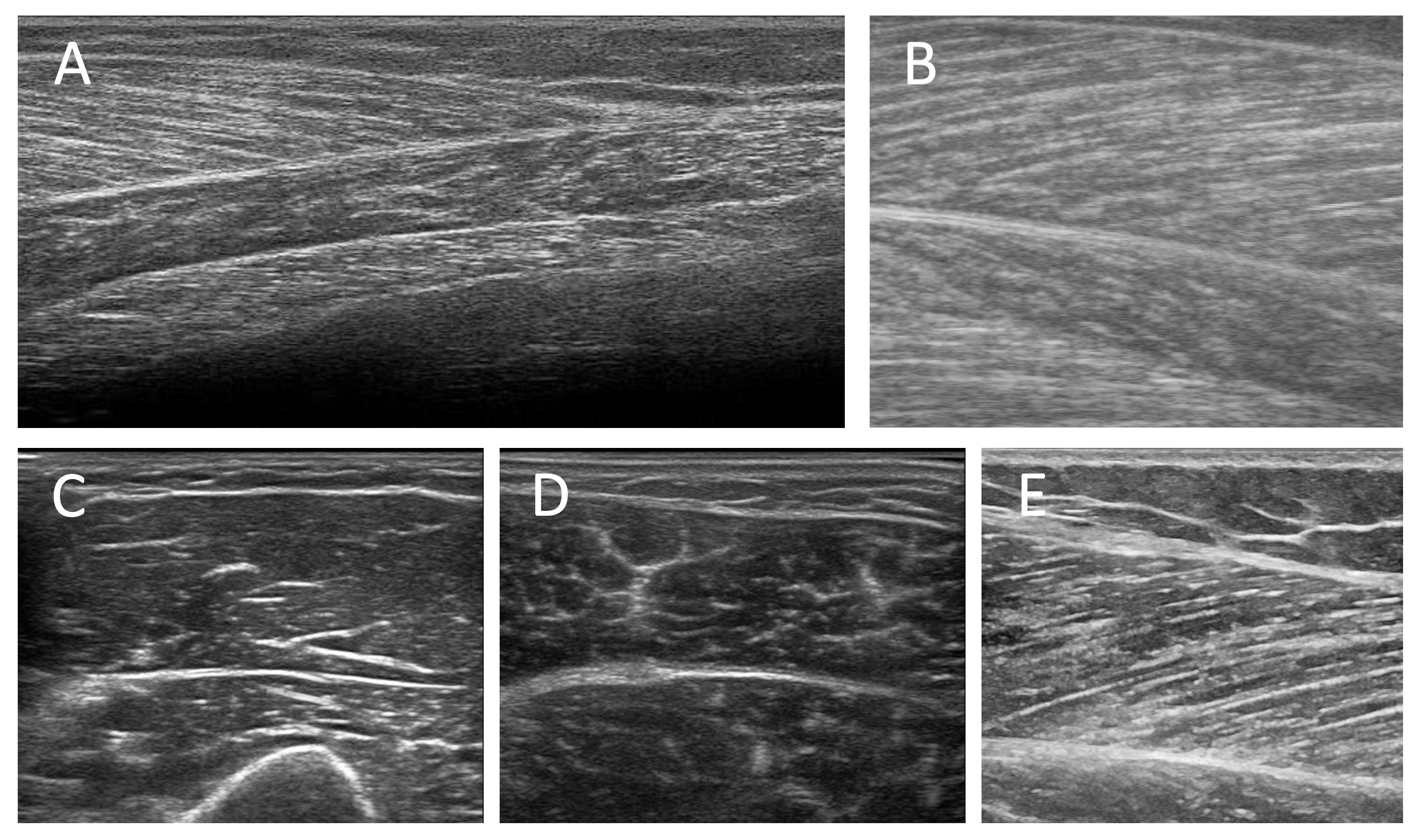}
    \caption{Example images from clinical datasets used to train the CycleGAN. Image A is of the gastrocnemius from dataset 1, image B is of the medial gastrocnemius and soleus from dataset 2, images C and D are of the biceps brachii and medial gastrocnemius respectively from dataset 3, and image E is of the vastus lateralis from dataset 4. All datasets are described in Table \ref{tab:1}.}
    \label{fig:2}
\end{figure}

Figure \ref{fig:2} displays example images from the clinical image repositories (Table \ref{tab:1}). As seen from the example images, these online datasets exhibit different contrasts, resolutions, and dynamic ranges. Additionally, the dataset is largely comprised of transverse views of muscle with 2,354 transverse muscle images as compared to 1,146 longitudinal images. 

Both the paired and unpaired datasets were split 80\%, 10\%, and 10\% into training, validation, and testing sets respectively. During training, images from both datasets were augmented by random flipping and cropping.

\begin{figure}[ht]
    \centering
    \includegraphics[width=0.5\textwidth]{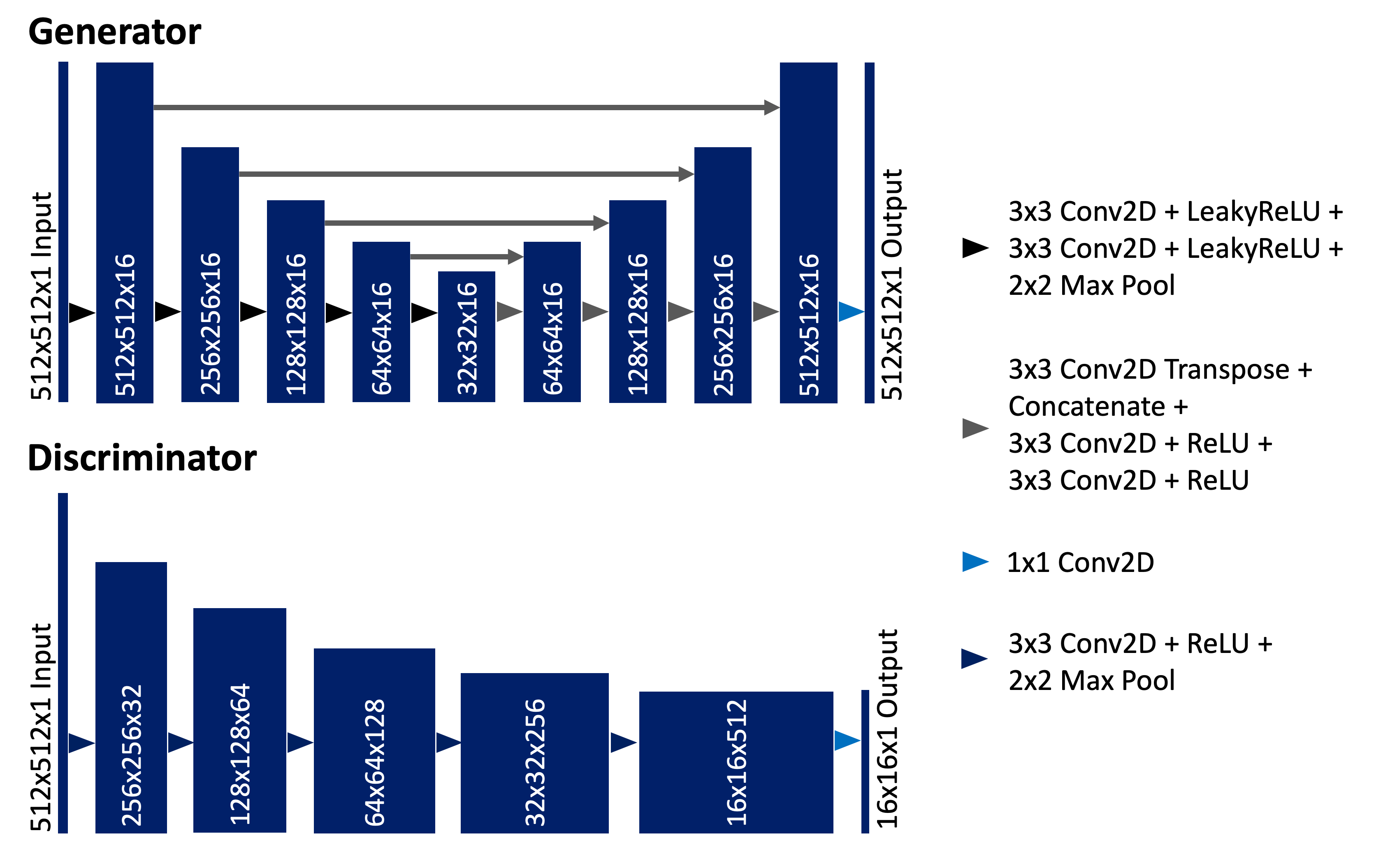}
    \caption{CycleGAN architecture. The first stage U-Net shares the same architecture as the CycleGAN’s generator.}
    \label{fig:3}
\end{figure}

\subsection{Machine Learning Models}
A two-stage machine learning model was trained on the paired and unpaired datasets. For the first stage, a U-Net was trained on the paired dataset to approximate not only plane wave compounding, but also a traditional image-processing pipeline consisting of histogram matching and unsharp filtering \cite{RN6}. The U-Net’s architecture is the same as the second stage CycleGAN’s generator presented in Figure \ref{fig:3}. This U-Net was trained over 100 epochs using a batch size of 1, the ADAM optimizer with a learning rate of 2e-4, and L1 loss. The training weights after the epoch with the lowest validation loss were saved.

The second stage uses a CycleGAN to translate the output of the first stage into images that resemble clinical muscle images. The CycleGAN consists of 2 generators and 2 discriminators responsible for performing style transfer between research scanner images and clinical scanner images \cite{RN31}. During training, the CycleGAN uses the unpaired dataset described above. During inference, the output of the first stage is used as the input for the forward generator of the CycleGAN. The CycleGAN is based upon MimickNet \cite{RN4} and is presented in Figure \ref{fig:3}. The output of the discriminator is a 16x16x1 matrix following the PatchGAN approach \cite{RN32}, and mean squared error was used for all loss functions following the LSGAN approach \cite{RN33}. The loss of the forward generator G ($\mathcal{L}_{gen}^G$), which translates from research scanner images to clinical scanner images, is described as follows:
\begin{equation}
    \begin{split}
        \mathcal{L}_{gen}^G = \lambda_G*\mathcal{L}_{GAN}^G 
        + \mathcal{L}_{cyc}^G + \mathcal{L}_{id}^G/2
    \end{split}
\end{equation}
Where $\mathcal{L}_{GAN}$ represents the adversarial loss, $\mathcal{L}_{cyc}$ represents the cycle-consistency loss, and $\mathcal{L}_{id}$ represents the identity loss. The loss of the reverse generator F ($\mathcal{L}_{gen}^F$), which translates from clinical scanner images to research scanner images, is similar:
\begin{equation}
    \begin{split}
        \mathcal{L}_{gen}^F = \lambda_F*\mathcal{L}_{GAN}^F 
        + \mathcal{L}_{cyc}^F + \mathcal{L}_{id}^F/2
    \end{split}
\end{equation}
Here, $\lambda_G$ and $\lambda_F$ are tunable hyperparameters.

To stabilize the training of the CycleGAN, parameter sweeps were performed over batch size, $\lambda_G$, $\lambda_F$, learning rate and learning rate schedule (using the ADAM optimizer), number of layers and filters (for both generators and both discriminators), and normalization techniques including batch, instance, and spectral normalization. The final parameters were a batch size of 1; a $\lambda_G$ of 1 and a $\lambda_F$ of 0.1; a piecewise learning rate of 1e-4, 5e-5, and 1e-5 after 0, 10,000, and 30,000 steps respectively; layers and filters as displayed in Figure \ref{fig:3}; and spectral normalization applied only to the discriminator of the research scanner image domain.

\subsection{Implementation in MATLAB and Verasonics}
As our Verasonics system was limited to MATLAB R2019a, it was necessary to import the stage 1 and stage 2 models, which were trained with Python 3.7 and TensorFlow 2.9.1, into MATLAB R2019a. To do so, the TensorFlow models were first converted into the Open Neural Network Exchange (ONNX opset-9) format and imported into MATLAB. Whereas TensorFlow’s implementation of the transposed convolution layer pads the input image, MATLAB’s implementation crops the output image. Due to this difference, these layers were replaced using MATLAB’s implementation while preserving the original kernel weights. Finally, the layers were assembled into MATLAB DAGNetworks. 

\begin{figure}[ht]
    \centering
    \includegraphics[width=0.5\textwidth]{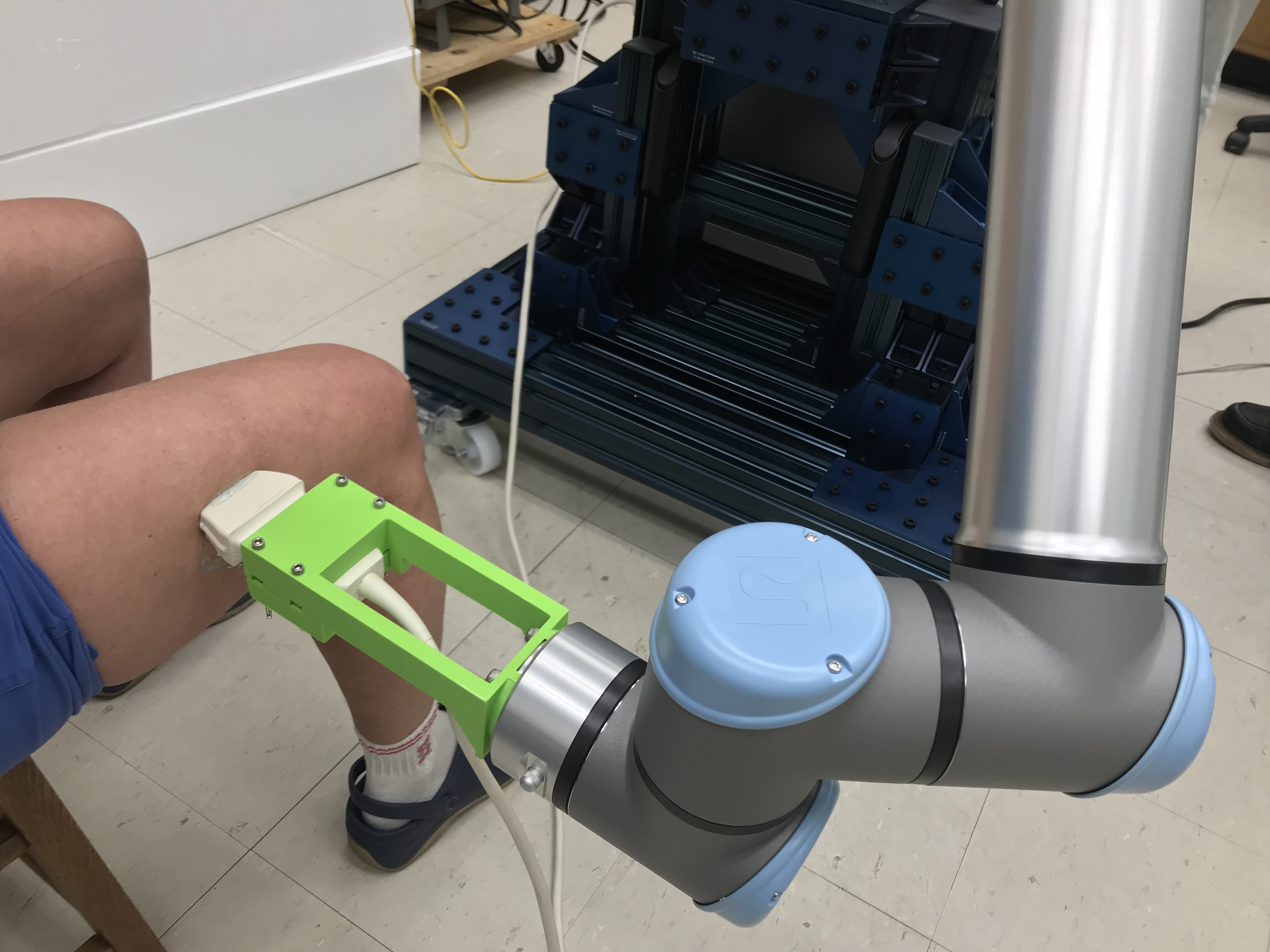}
    \caption{Setup used to image a healthy volunteer using the Verasonics Vantage™ system and L7-4 transducer.}
    \label{fig:4}
\end{figure}

To incorporate the model into the Verasonics acquisition sequences, a Verasonics External Process object was created which resizes, envelope-detects, and log-compresses the plane wave data before post-processing the beamformed image with the DAGNetworks. The output is displayed on a persistent MATLAB figure which serves as a custom display window. The custom display window displays the plane wave image produced by the Verasonics and the sequential outputs of the first and second stages in real-time. Using single plane wave transmits and dynamic receive focusing with an F-number of 0.81, this system was used to image the vastus lateralis of a healthy volunteer as seen in Figure \ref{fig:4}.

For real-world frame rate comparisons, the External Process object was modified to either use histogram matching (to serve as a baseline reference), the first ML stage, or the combined ML stages, and the average frame rate was evaluated with each method.

\begin{figure}[ht]
    \centering
    \includegraphics[width=0.5\textwidth]{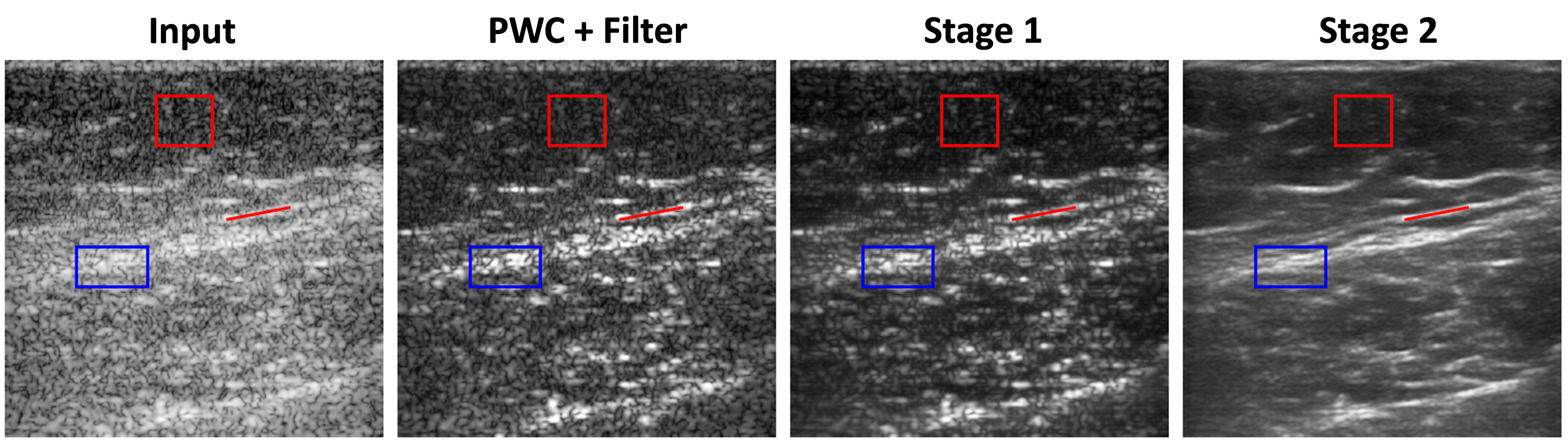}
    \caption{Example of ROIs used to calculate image metrics. Standard deviation was calculated over the red box to quantify speckle. CNR was calculated over the red and blue boxes. Standard deviation was calculated over the red line segment to characterize fiber and fascicle cohesiveness.}
    \label{fig:5}
\end{figure}

\subsection{Reader study and Analysis}
Due to the difficulty of obtaining accurate quantitative measurements of image quality, especially for unpaired images as in the case of the CycleGAN, a reader study was performed by two neurologists from the Duke University Medical Center (LHW and YH). Each reader evaluated 24 sets of muscle ultrasound images. This sample size was determined using a power analysis with an estimated Cohen’s effect size of 0.35 \cite{RN34}. Each set was composed of 4 images: the plane wave input image, the plane wave compounded and filtered image, the output of the first ML stage, and the output of the second ML stage. Readers were asked to score each image based on 2 criteria on a Likert-type scale describing subjective levels of image quality from 0 to 3. The criteria are described as follows:
\begin{enumerate}
    \item Amount of speckle (or noise) present in the image. A score of 0 corresponds to low amounts of speckle in the image, and a score of 3 corresponds to large amounts of speckle.
    \item Structural fidelity (how clear and distinct the muscle fibers and fascicles are). A score of 0 corresponds to low structural fidelity, and a score of 3 corresponds to high structural fidelity.
\end{enumerate}
Readers were encouraged to use the full range of the scoring scale (from 0 to 3). To help familiarize readers with the image presentation format and the range of image qualities in the dataset, the first 4 sets of images were used as examples and were excluded during data analysis.

In addition to the reader study, ROIs were applied to the same 24 sets of muscle ultrasound images and evaluated with imaging metrics. To evaluate speckle, standard deviation was calculated on a rectangular ROI placed over a hypoechoic region of the image. Contrast-to-noise ratio (CNR) was calculated using rectangular ROIs placed over hyperechoic and hypoechoic regions of the image. The following equation was used for CNR:
\begin{equation}
CNR=\frac{\mu_2-\mu_1}{(\sigma_1+\sigma_2)/2}
\end{equation}
To evaluate fiber cohesiveness, standard deviation was calculated over a line segment placed on a muscle fiber or fascicle. An example set of images with ROIs applied is shown in Figure \ref{fig:5}.

\section{Results}
\subsection{Stage 1}
\begin{figure}[ht]
    \centering
    \includegraphics[width=0.5\textwidth]{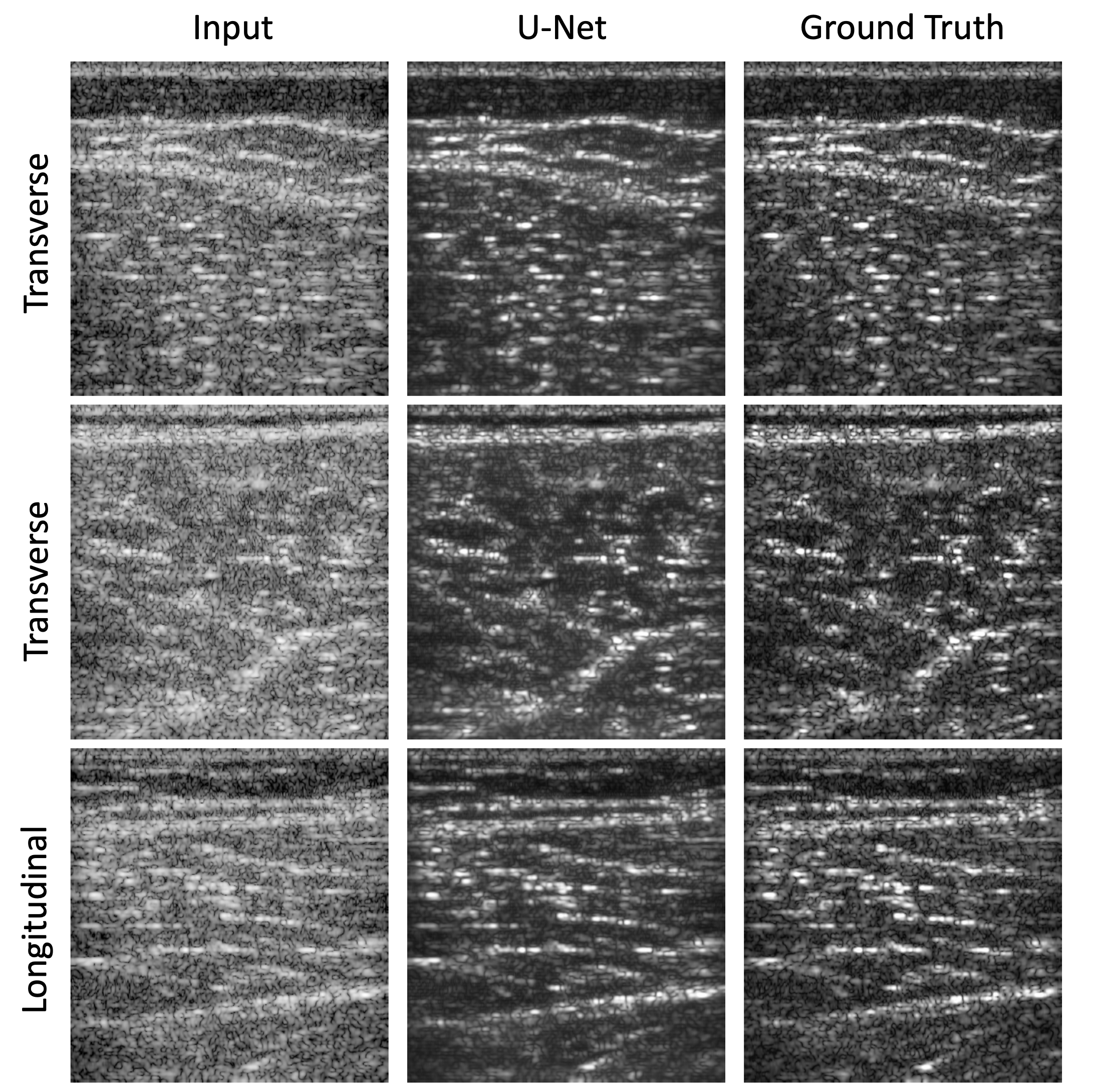}
    \caption{Example images taken from the test dataset. The first column shows the single plane wave images used as the input for the U-Net, and the second column displays the outputs of the U-Net. The third column displays the ground truth images. The first two rows display transverse images, while the third row displays longitudinal images of muscle.}
    \label{fig:6}
\end{figure}

As seen in Figure \ref{fig:6}, the first stage U-Net was able to produce images that closely resemble the ground truth produced by PWC and filtering using a traditional image processing pipeline (histogram matching and unsharp filtering). To compare the U-Net with the traditional image-processing pipeline, normalized RMSE, structural similarity index measure (SSIM), and processing time per image are tabulated in Table \ref{tab:2}. The normalized RMSE and SSIM for the plane wave input and the U-Net were calculated against the ground truth. The RMSE was normalized using the L2 norm of the ground truth image, and a sliding window size of 11x11 pixels was used for the SSIM. As expected, the U-Net had a lower average normalized RMSE of 0.274 and a higher average SSIM of 0.622 when compared to the plane wave input. 

One of the advantages of machine learning models is their facile implementation on GPUs which greatly accelerate training and inference times \cite{RN35}. Traditional image processing libraries often lack GPU support resulting in slower computation times. The U-Net had an average inference time of 18 ± 2 milliseconds when running on a Tesla V100. In contrast, the traditional image processing pipeline had an average runtime of 202 ± 18 milliseconds per image while running on an Intel Xeon Gold 6252 CPU (2.10 GHz).

\begin{table}[ht]
    \centering
    \caption{Comparisons for the first stage. The original plane wave input and the U-Net output are compared against the PWC and filtered ground truth using two metrics: normalized RMSE (normalized with the Euclidean norm of the ground truth) and SSIM. Per-image inference times using the U-Net were measured on both CPU and GPU. The time to filter the PWC image using the traditional image processing pipeline (histogram matching and unsharp filtering) was measured on the CPU. Values are displayed as metric/time ± standard deviation.}
    \label{tab:2}
    \begin{tabular}{>{\raggedleft\arraybackslash}m{.19\textwidth}*{3}{>{\centering\arraybackslash}m{.065\textwidth}}}
        \toprule
        & \textbf{Plane wave (input)} & \textbf{U-Net} & \textbf{Filtering the PWC image} \\
        \midrule
        \textbf{Average normalized RMSE} & 0.547 ± 0.140 & 0.274 ± 0.023 & N/A \\
        \midrule
        \textbf{Average SSIM} & 0.344 ± 0.161 & 0.622 ± 0.041 & N/A \\
        \midrule
        \textbf{Average time per image on CPU (ms)} Intel Xeon Gold 6252 CPU at 2.10 GHz & N/A & 350 ± 42 & 202 ± 18 \\
        \midrule
        \textbf{Average time per image on GPU (ms)} Tesla V100-PCIE-16GB & N/A & 18 ± 2 & N/A \\
        \bottomrule
    \end{tabular}
\end{table}

\subsection{Stage 2}
\begin{figure*}[ht]
    \centering
    \includegraphics[width=\textwidth]{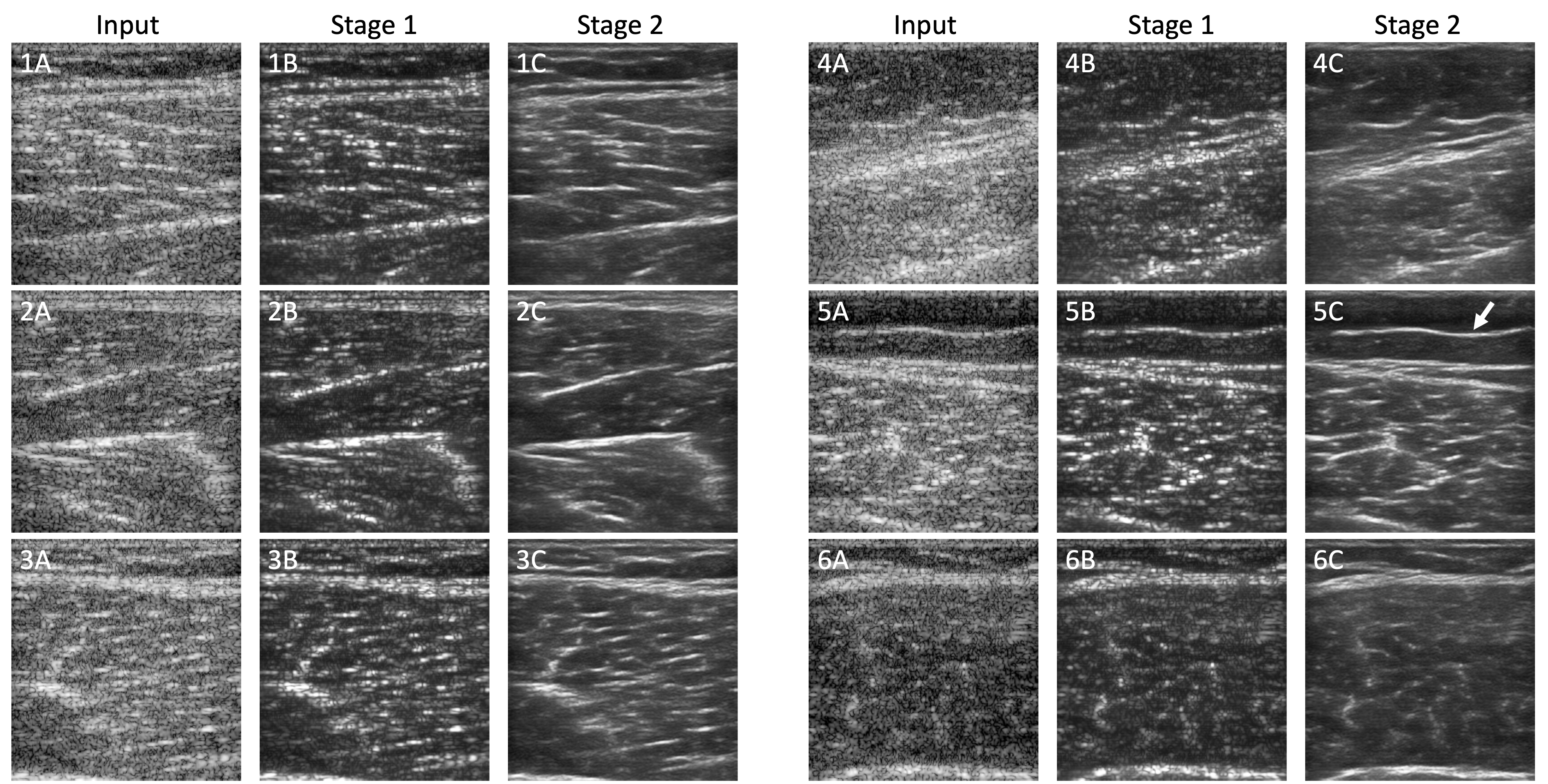}
    \caption{Images from the test dataset following processing by the first-stage U-Net and the second-stage CycleGAN. Rows 1, 2, and 3 represent longitudinal images of muscle, while rows 4, 5, and 6 represent transverse images. Note the sharpened hyperechoic line in 5C marked by an arrow.}
    \label{fig:7}
\end{figure*}

Using the output of the first stage U-Net as the input to the second stage CycleGAN, Figure \ref{fig:7} shows outputs of the CycleGAN during inference.  The model increased mean CNR from 3.67 in the input images to 5.11 as seen in Table \ref{tab:3}. The model also decreased the amount of speckle present in the images which is desirable in clinical images \cite{RN5}, reducing the standard deviation of the speckle from an average of 0.208 in the input images to 0.068 (Table \ref{tab:3}). The model also improved the cohesiveness of the muscle fibers and fascicles by connecting the fibers and fascicles that were interrupted by the speckle pattern. Mean fiber/fascicle standard deviation decreased from 0.162 to 0.115 (Table \ref{tab:3}). This sharpening can be clearly seen in Figure \ref{fig:7}, image 5C, where the hyperechoic line clearly bisects the fatty tissue.

\begin{table}[ht]
    \centering
    \caption{Image metrics (mean ± standard deviation) calculated over 24 sets of images. Figure \ref{fig:5} displays example ROIs used to calculate these metrics.}
    \label{tab:3}
    \begin{tabular}{>{\raggedleft\arraybackslash}m{.1\textwidth}*{3}{>{\centering\arraybackslash}m{.09\textwidth}}}
        \toprule
        & \textbf{Speckle Std Dev} & \textbf{CNR} & \textbf{Fiber Std Dev} \\
        \midrule
        \textbf{Input} & 0.208 ± 0.018 & 3.67 ± 1.09 & 0.162 ± 0.042 \\
        \textbf{PWC + Filter} & 0.139 ± 0.023 & 3.37 ± 0.899 & 0.274 ± 0.097 \\
        \textbf{Stage 1} & 0.099 ± 0.020 & 3.94 ± 0.943 & 0.241 ± 0.069 \\
        \textbf{Stage 2} & 0.068 ± 0.014 & 5.11 ± 1.58 & 0.115 ± 0.043 \\
        \bottomrule
    \end{tabular}
\end{table}

\subsection{Reader Study}
\begin{figure}[ht]
    \centering
    \includegraphics[width=0.5\textwidth]{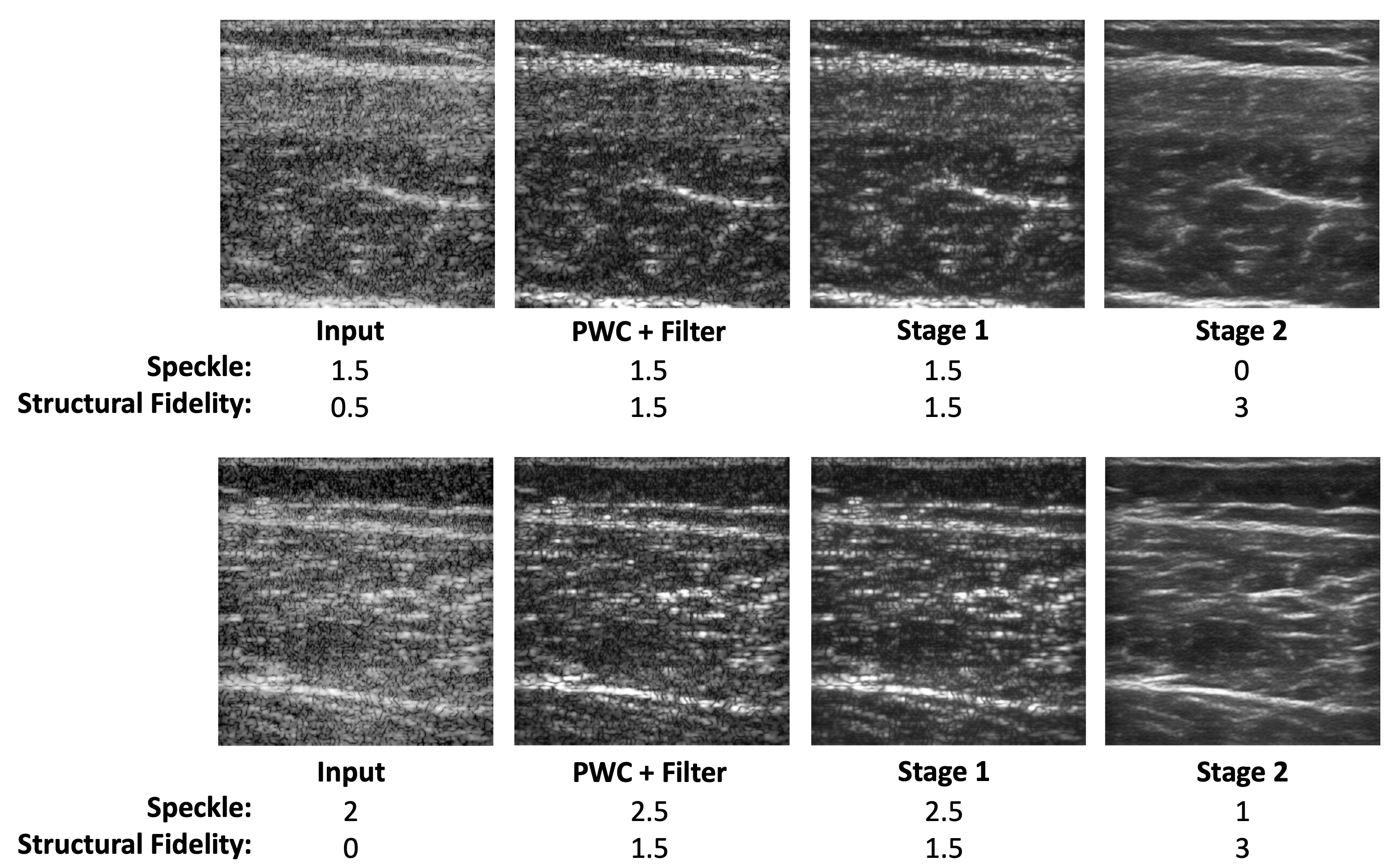}
    \caption{Two sets of paired images evaluated by the readers. Image order was randomized for the reader study. Here, labels are provided below each image (Input, PWC + Filter, Stage 1, or Stage 2). Scores for speckle and structural fidelity, averaged between the two readers, are also provided.}
    \label{fig:8}
\end{figure}

\begin{figure}[ht]
    \centering
    \includegraphics[width=0.5\textwidth]{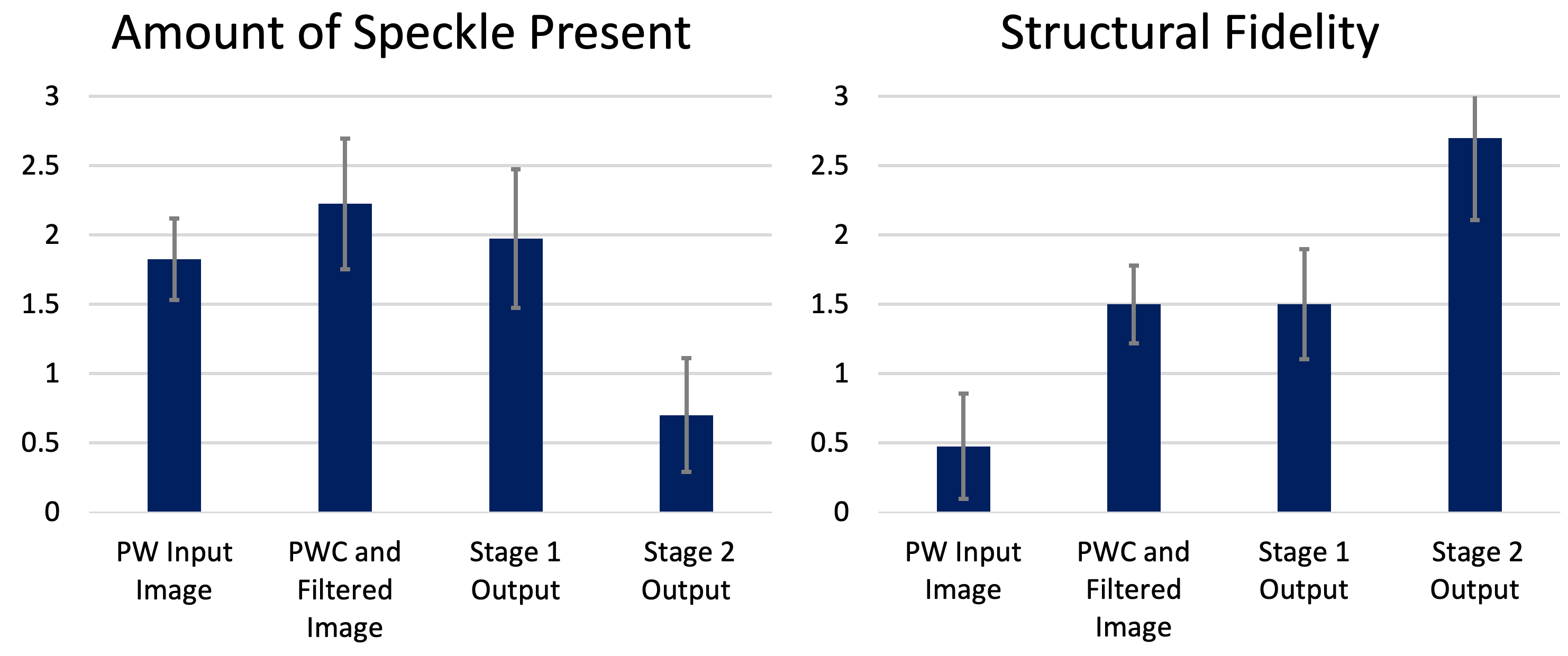}
    \caption{Average scores and standard deviations (sample size of 20 with 2 readers) of the speckle and structural fidelity metrics.}
    \label{fig:9}
\end{figure}

The speckle and structural fidelity scores for each image were averaged between the two readers. Figure \ref{fig:8} provides 2 examples of the paired images that the readers evaluated and their corresponding scores. The average scores and standard deviations of the entire reader study are presented in Figure \ref{fig:9}. A Friedman’s ANOVA was conducted for each quality metric between four groups: the input image, the PWC and filtered image, the output of the stage 1 U-Net, and the output of the stage 2 CycleGAN. Following the Friedman’s ANOVA, which resulted in significant p-values $<$0.01 for both the speckle and structural fidelity metrics, a post-hoc Nemenyi’s test was performed. For the speckle quality metric, scores for the plane wave input, PWC and filtered, and stage 1 images did not differ significantly. However, the speckle score for the stage 2 CycleGAN was significantly lower than the other three groups. For the structural fidelity metric, the scores for the PWC and filtered images and the stage 1 images were both 1.5. However, structural fidelity scores for all other pairs of groups differed significantly. Table \ref{tab:4} shows the p-values of the post-hoc Nemenyi’s test.

These results show that the two-stage machine learning model successfully decreased the amount of speckle present in the plane wave images while improving their structural fidelity. However, the PWC and filtering process (as well as the first stage) increased the amount of speckle present in the images as rated by the readers. There were no significant differences between scores for the stage 1 images and PWC and filtered images, providing further support that the first stage machine learning model successfully emulated the traditional PWC and filtering image-processing pipeline.

\begin{table}[ht]
    \centering
    \caption{P-values of the post-hoc Nemenyi’s test.}
    \label{tab:4}
    \begin{tabular}{*{2}{>{\centering\arraybackslash}m{.12\textwidth}}*{2}{>{\centering\arraybackslash}m{.07\textwidth}}}
        \toprule
        \textbf{Group 1} & \textbf{Group 2} & \textbf{Speckle P-values} & \textbf{Structural Fidelity P-values} \\
        \midrule
        PW Input & PWC and Filtered & 0.092 & $<$0.05 \\
        PW Input & Stage 1 & 0.93 & $<$0.05 \\
        PW Input & Stage 2 & $<$0.05 & $<$0.05 \\
        PWC and Filtered & Stage 1 & 0.32 & 1.0 \\
        PWC and Filtered & Stage 2 & $<$0.05 & $<$0.05 \\
        Stage 1 & Stage 2 & $<$0.05 & $<$0.05 \\
        \bottomrule
    \end{tabular}
\end{table}

\subsection{Verasonics}
\begin{figure}[ht]
    \centering
    \includegraphics[width=0.5\textwidth]{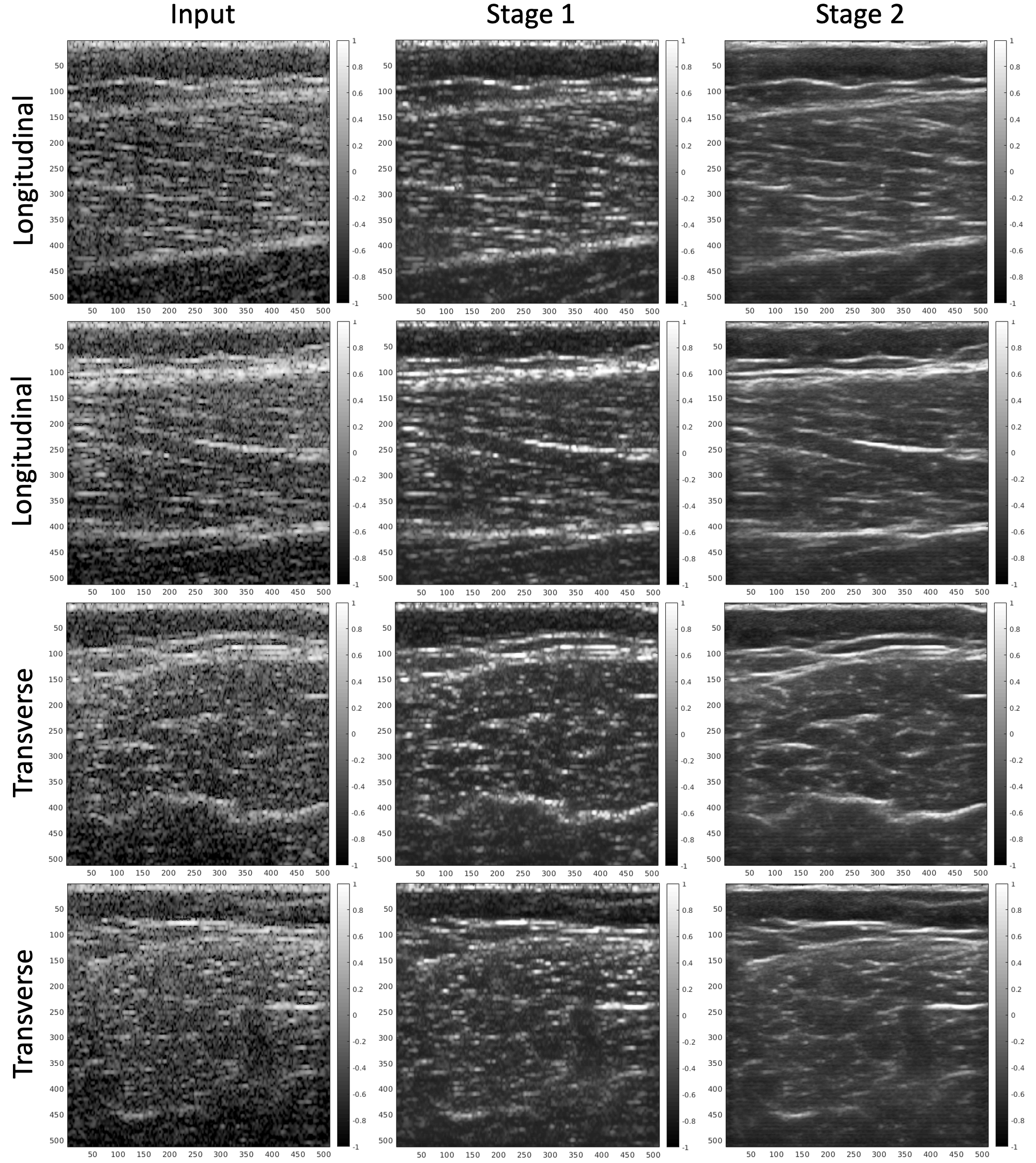}
    \caption{Real-time images of the vastus lateralis were acquired from a healthy volunteer. The first column shows the beamformed, resized, envelope-detected, and log-compressed image produced by the Verasonics. This image is then sequentially processed by the first stage (column 2) and second stage (column 3) in real-time. The first two rows show screenshots taken during along-the-fiber (longitudinal) imaging, and the last two rows show screenshots taken during across-the-fiber (transverse) imaging.}
    \label{fig:10}
\end{figure}

Screenshots of the custom display window during real-time imaging with the Verasonics are shown in Figure \ref{fig:10}, where each row of images is a separate screenshot. Table \ref{tab:5} demonstrates that histogram matching had the highest frame rate of 81.8 ± 1.2 FPS and processing with the combined stages had the slowest frame rate of 28.5 ± 0.6 FPS.

\begin{table}[ht]
    \centering
    \caption{Average frame rates and standard deviations on the Verasonics Vantage™ system when post-processing with the first stage and combined stages (stages 1 and 2). Post-processing with histogram matching (without unsharp filtering) is included as a baseline reference.}
    \label{tab:5}
    \begin{tabular}{ccc}
        \toprule
        \textbf{Histogram Matching} & \textbf{Stage 1} & \textbf{Stage 1} \\
        \midrule
        81.8 ± 1.2 FPS & 40.0 ± 1.8 FPS & 28.5 ± 0.6 FPS \\
        \bottomrule
    \end{tabular}
\end{table}

\section{Discussion}
The first stage ML model was trained to emulate plane wave compounding and a traditional image processing pipeline. As seen in Figure \ref{fig:6} and Table \ref{tab:2}, not only did it successfully emulate the traditional image-processing pipeline by halving the normalized RMSE and nearly doubling the SSIM as compared to the input plane wave image, the first stage ML model was able to do so in less time. However, the SSIM of 0.622 for the U-Net is still a relatively poor score for the SSIM metric \cite{RN36}. Additionally, readers rated the traditional image processing pipeline and first stage as having increased amounts of speckle (Figure \ref{fig:9}), likely due to the unsharp filter amplifying the speckle. 
  
The second stage CycleGAN generates images that closely resemble those produced by clinical scanners as seen in Figure \ref{fig:7}. Combining the two stages together, the combined ML model was able to process plane wave images of skeletal muscle and generate clinical style images in real-time at a frame rate of 28.5 ± 0.6 FPS (Table \ref{tab:5}). This frame rate can easily be considered “real-time,” demonstrating the utility of using machine learning models for real-time ultrasound enhancement. As seen in Table \ref{tab:3}, the model reduced speckle while improving CNR and fiber cohesiveness. This improvement in image quality is especially noticeable in Figure \ref{fig:10} where the input image is heavily corrupted by speckle and noise, but the combined model was able to connect muscle fibers and fascicles while suppressing speckle.

One limitation of the model may be attributed to the unpaired dataset. As seen in Table \ref{tab:1}, there are nearly twice the number of transverse images of muscle as there are longitudinal. This imbalance may bias the CycleGAN to produce images that resemble transverse images of muscle even when the input image is a longitudinal image of muscle. Additionally, CycleGANs are used to perform style transfer from one style to another \cite{RN31}, and the heterogeneous contrasts, resolutions, and dynamic ranges between the four online repositories of clinical images may pose challenges for the CycleGAN when emulating a clinical style.

The use of generative machine learning models in medical imaging is also limited due to their black box nature \cite{RN37} and ability to hallucinate and introduce false information into the data \cite{RN38}. The stage 2 CycleGAN, which sharpens and connects muscle fibers and fascia, can also generate structures that were not originally present. The two-stage design of the model helps alleviate these limitations by enabling the reader to select between a highly processed clinical-style image that may contain hallucinated structures, or a less-processed image that is more likely to preserve the original information in the image.

Enhancing plane wave images of muscle to improve muscle and fiber definition is useful because during SWEI acquisitions, large numbers of plane wave images are acquired. While the ML-based image enhancements developed herein are not applicable to SWEI tracking data, enhanced fiber and fascia definition can be helpful for later analyses such as muscle segmentation and determining muscle fiber orientation in B-mode images \cite{RN28, RN39, RN40}. Additionally, one significant advantage of this approach is that it can help streamline SWEI acquisitions on the Verasonics. During SWEI acquisitions, it is helpful to periodically obtain B-mode images to verify that the probe is accurately positioned and is imaging the correct target. However, adding B-mode sequences to SWEI acquisition sequences adds complexity and interrupts the SWEI acquisition. Instead, plane wave images acquired during SWEI can be continuously processed and displayed using this model without interruption, providing real-time feedback during acquisitions and thus improving data quality. 

\subsection{Future Work}
To improve CycleGAN performance, especially for longitudinal muscle images, it would be prudent to obtain a high-quality clinical dataset with a balanced number of longitudinal and transverse images. A clinical dataset with uniform contrast, dynamic range, and resolution may also facilitate training of the CycleGAN and improve model performance. 

This two-stage model is very application-specific. While the unpaired clinical dataset consists of a variety of clinical repositories, the paired research dataset consists only of plane wave images of the vastus lateralis acquired on a single probe (L7-4) and acquisition sequence. Training the CycleGAN using data acquired with multiple different transducers, acquisition sequences, and muscle targets may lead to a more generalizable model. The current acquisition sequence reconstructs 97x191 pixel images, and the architecture of the U-Net model restricts the input and output image size to a relatively small resolution of 512x512. In future work, the acquisition sequence and U-Net model can be modified and expanded to generate higher-resolution images. With the success of diffusion models in image generation, cycle-consistent diffusion models can also be implemented for unpaired ultrasound image translation \cite{RN41, RN42}. However, the iterative nature of diffusion models results in long sampling times that limit their ability to perform real-time inference. While methods such as strided sampling and progressive distillation have been developed to speed up inference of diffusion models \cite{RN43, RN44}, these methods are generally still slower than the single-step inference of GANs. 

\section{Conclusions}
This project demonstrates the utility of using a two-stage machine learning model to enhance ultrasound images of skeletal muscle in real-time at a frame rate of 28.5 ± 0.6 FPS, specifically muscle images acquired using single plane waves on research scanners. The project also describes a method for implementing TensorFlow models in the Verasonics for real-time image processing. The 2-stage model decreased speckle, as quantified by standard deviation, from 0.208 ± 0.018 in the input images to 0.068 ± 0.014. The model also increased CNR from 3.67 ± 1.09 to 5.11 ± 1.58. Fiber cohesiveness, as quantified by standard deviation across a line fiber or segment, also decreased from 0.162 ± 0.042 to 0.115 ± 0.043. Finally, readers from the Duke University Medical Center rated the images generated by the model as having significantly lower speckle content and higher structural fidelity. Future work may explore expanding the training dataset for improved model performance and generalizability.

\section*{Code Availability}
Code for this paper, including TensorFlow models, training, and detailed Verasonics implementation, is available at  \href{https://github.com/reedchen19/cyclegan_verasonics}{https://github.com/reedchen19/cyclegan\_verasonics}

\section*{Acknowledgments}
This work was supported by the Duke University Pratt Research Fellows Program.

\bibliographystyle{abbrv}
\bibliography{references}

\end{document}